%% file: main.tex
\documentclass[10pt,conference]{IEEEtran}
\usepackage[normalem]{ulem}
\usepackage[hyphens]{url}
\usepackage[sort,nocompress]{cite}
\usepackage[bookmarks=true,breaklinks=true,letterpaper=true,colorlinks,citecolor=blue,linkcolor=blue,urlcolor=blue]{hyperref}
\usepackage{fancyhdr}
\usepackage{amsmath,amssymb,amsfonts}

\usepackage{tikz}
\usepackage{colortbl}
\usetikzlibrary{positioning}

\usepackage{makecell}

\usepackage{amsmath,amssymb}
\usepackage{array}
\usepackage{xcolor}

\usepackage{algorithmic}
\usepackage{graphicx}
\usepackage{textcomp}
\usepackage{xcolor}
\usepackage{pifont}
\usepackage{colortbl}
\usepackage{booktabs}
\usepackage[hyphens]{url}
\usepackage{enumitem}
\usepackage{caption}
\usepackage{subcaption}
\usepackage{pifont}

\definecolor{deepgreen}{RGB}{34,139,34} 
\definecolor{deepyellow}{RGB}{218,165,32} 

\usepackage{tikz}

\newcommand*\circled[1]{\tikz[baseline=(char.base)]{
            \node[shape=circle,fill,inner sep=0.5pt] (char) {\textcolor{white}{#1}};}}

\newcommand{\cmark}{\textcolor{green!70!black}{\checkmark}}
\newcommand{\xmark}{\textcolor{red!80!black}{\ding{55}}}

\def\BibTeX{{\rm B\kern-.05em{\sc i\kern-.025em b}\kern-.08em
    T\kern-.1667em\lower.7ex\hbox{E}\kern-.125emX}}

\definecolor{Che}{RGB}{0,128,128}
\definecolor{Geng}{RGB}{128,0,128}
\newcommand{\HPCA}[1]{\textcolor{black}{#1}}

\title{Secure Scattered Memory: Rethinking Secure Enclave Memory with Secret Sharing}

\author{
\IEEEauthorblockN{
Haoran Geng$^{*}$, Yuezhi Che$^{\dagger}$, 
Dazhao Chen$^{\dagger}$, Michael Niemier$^{*}$, 
Xiaobo Sharon Hu$^{*}$
}
\IEEEauthorblockA{
$^{*}$Department of Computer Science and Engineering, 
University of Notre Dame, Notre Dame, IN, USA, 46556 \\
Email: hgeng@nd.edu, mniemier@nd.edu, shu@nd.edu
}
\IEEEauthorblockA{
$^{\dagger}$Department of Computer Science, 
Wuhan University, Wuhan, China \\
Email: cheyuezhi@whu.edu.cn, dcheng@whu.edu.cn
}
}

\begin{document}
\maketitle
\thispagestyle{plain}
\pagestyle{plain}


\begin{abstract}

The rise of cloud computing demands secure memory systems that ensure data confidentiality, integrity, and freshness against replay attacks. Existing schemes such as AES-XTS, AES-GCM, and AES-CTR each trade performance for security, with only AES-CTR plus Message Authentication Codes (MAC) and Merkle Trees (MT) providing full protection—at the cost of substantial counter and MT overhead. This paper introduces Secure Scattered Memory (SSM), a novel scheme that replaces counter-based encryption with polynomial-based secret sharing. Each data block is encoded into multiple cryptographically independent shares distributed across memory, inherently preventing information leakage while ensuring integrity and freshness through mathematical reconstruction properties. Implemented and synthesized in a 28 nm commercial PDK, SSM occupies 0.27 mm² and consumes 284.53 mW. Experiments show only 10 \% and 8 \% performance overhead over AES-XTS and AES-GCM, respectively, while outperforming Morphable Counter (MICRO 2018) by up to 40\%, achieving 12\% better performance than EMCC/RMCC (MICRO 2022), and exceeding COSMOS (MICRO 2025) by 3\%.

\end{abstract}
\input{introduction-new}

\input{02_Background}

\input{03_Motivation}

\input{04_Secure_Scattered_Memory}

\input{05_Threat_Model}
\input{07_Methology}

\input{07_Experimental_Results}

\input{08_Related_Work}

\input{09_Conclusion}


\bibliographystyle{IEEEtranS}
\bibliography{refs}

\end{document}

%% file: introduction-new.tex
\section{Introduction}
\label{sec:introduction}
Securing sensitive and private data is crucial as companies migrate computational workloads to the cloud for cost savings, scalability, and flexibility. However, this raises significant security concerns. When data is stored and processed on remote servers, owners relinquish direct control over physical infrastructure, making it susceptible to unauthorized access and breaches by compromised or malicious servers—a critical risk that must be addressed.


To address security challenges, modern processors integrate hardware-based memory encryption for data confidentiality, integrity, and freshness. Intel SGX employs AES-CTR with MAC and MT (SGXv1~\cite{SGX,SGX_explain,sgx_suvery}) for strong security, including replay resistance, but incurs substantial performance and storage overhead from frequent counter updates and MT traversals. Performance-optimized alternatives like AES-XTS and AES-GCM have been widely adopted to mitigate this. AES-XTS is used in Intel SGXv2~\cite{sgx_v2,sgx_v2_1} and AMD SEV-family architectures (SEV, SEV-ES, SEV-SNP)~\cite{amd_sev_1,amd_sev_2,amd_white_paper}, while AES-GCM is deployed in DDR5 modules from Rambus~\cite{rambus_1,rambus_2} and emerging standards like CXL.Mem~\cite{cxl_1,cxl_2,cxl_3}.

While these schemes offer reasonable trade-offs between performance and protection, both AES-XTS and AES-GCM intentionally relax certain security guarantees in favor of simplified implementation and lower overhead. In particular, AES-XTS lacks any form of cryptographic integrity protection and does not enforce data freshness, making it inherently vulnerable to substitution and replay attacks~\cite{XTS_sca1,XTS_sca2,NIST_XTS}; AES-GCM, while offering stronger integrity guarantees through authenticated encryption, similarly omits version number (VN) updates or monotonic counters, and therefore remains susceptible to replay attacks under adversarial conditions~\cite{GCM_WEAK1,GCM_WEAK2,synergy,NIST_GCM}. Among these schemes, SGXv1-like provides the most comprehensive protection, including confidentiality, integrity, and replay attacks~\cite {mgx,softvn}.

However, SGXv1-like incurs significant performance overhead due to three main factors: 
\circled{1} \textbf{VN updates}: Each encryption or decryption operation requires accessing and updating the VN values, which causes an increase in memory bandwidth consumption due to the need to read and write VN blocks \cite{emcc,rmcc,mgx,softvn,osiris,horus}. 
\circled{2} \textbf{Integrity checks}: The system must verify data integrity upon each read operation by recomputing and checking the MAC. Additionally, using MT to maintain the integrity of VN blocks compounds this overhead. Modifications to VN blocks require updates to the MT, involving read and write operations on tree nodes, which significantly increase memory traffic.

To address performance issues, VNs are cached in a  cache within the memory controller. However, this cache exhibits poor performance even with recent optimizations \cite{countercache,countercache_2,countercache_3,CCmiss_1,CCmiss_2,CCmiss_3,CCmiss_4,CCmiss_5} because it stores VNs for data blocks that have already missed in the LLC—precisely those with poor locality. State-of-the-art optimizations attempt to modify the architecture to increase VN cache hit rates \cite{emcc, rmcc}, but VN misses still incur high overhead and introduce additional hardware complexity. Although approaches like MGX \cite{mgx} and SoftVN \cite{softvn} eliminate MT structures by using specialized hardware or software to ensure freshness and uniqueness, they require substantial system modifications that raise compatibility concerns and introduce potential security vulnerabilities. Other approaches \cite{counterlight} attempt to combine SGXv1-like protection with AES-XTS, but this hybrid design does not fully protect replay attacks.

In this paper, we rethink secure memory design from a fundamentally different perspective. We propose Secure Scattered Memory (SSM), a novel memory protection scheme that diverges from traditional encryption methods. SSM leverages the concept of secure sharing to safeguard both data confidentiality and integrity. The main innovation of SSM lies in decomposing the original data into multiple secret shares rather than encrypting the data directly.  
These shares are scattered across memory and are individually meaningless. In SSM, no single data block reveals significant information; only the complete assembly of the required shares can reconstruct the original data. Data integrity is inherently ensured through the reconstruction process: the original data can only be accurately reconstructed when all required shares are assembled, thereby confirming that each share remains unaltered.

To handle replay attacks, SSM incorporates a dynamic share regeneration process. With each write operation, new shares are generated uniquely and incrementally, which mimics the VN updates in the AES-CTR encryption. This dynamic adjustment ensures data freshness and provides robust protection against replay attacks. By adopting a novel approach to data confidentiality and integrity and eliminating the need for traditional MT structures, SSM enhances the memory protection security framework while addressing the performance overheads associated with conventional methods.

The main contributions of SSM are highlighted below:

\begin{itemize}
    \item We analyze the overheads of existing secure memory schemes, emphasizing inefficiencies from VN updates and integrity verification.
    \item We propose SSM, an efficient memory protection scheme without using conventional encryption methods, that ensures data confidentiality, verifies integrity, and prevents replay attacks.
    \item  We implement and synthesize SSM's hardware using Cadence Genus with a 28nm commercial PDK, demonstrating practical hardware feasibility and providing detailed area, power, and performance metrics (0.270 mm², 284.53 mW at 28nm).
\end{itemize}

Our experimental results show that SSM introduces only 10\% and 8\% performance overhead compared to AES-XTS and AES-GCM, respectively, while providing protection against replay attacks. Moreover, SSM achieves up to 40\% higher performance overhead compared to state-of-the-art AES-CTR-based designs.


%% file: 02_Background.tex
\section{Background}
\label{sec:background}

\subsection{Memory Encryption} \label{bg:ms}

\subsubsection{AES-XTS and AES-GCM}

\textbf{AES-XTS}~\cite{XTS_sca1,XTS_sca2,NIST_XTS} is a block-cipher mode originally designed for block-oriented storage and later adopted in Intel SGX v2, TDX, and AMD SEV. It employs two AES keys—one to derive a location-dependent \emph{tweak} from the logical address and another to encrypt the tweaked plaintext—so that identical data at different addresses yield distinct ciphertexts. XTS provides efficient confidentiality without relying on per-block counters or version numbers, but it lacks built-in authentication and replay resistance, leaving systems vulnerable to tampering or rollback attacks.

\textbf{AES-GCM}~\cite{GCM_WEAK1,GCM_WEAK2,synergy,NIST_GCM}, used in CXL.Mem and Rambus DDR5, extends security by integrating confidentiality and integrity in a single pass. It encrypts data in counter mode while generating a Galois-field authentication tag over ciphertext and associated metadata. This enables high throughput and low latency but requires careful nonce management—reusing or losing synchronization of a nonce–key pair compromises both confidentiality and authenticity. As a result, despite its popularity in network protocols, GCM sees limited adoption in large-scale secure memory, where nonce tracking and replay prevention across vast address spaces are difficult to guarantee.

\subsubsection{SGXv1-like}

In SGXv1-like schemes, each 64-byte data block pairs with a unique 64-bit VN for encryption and integrity protection. While standard implementations group 8 VNs into a counter block, optimizations like Morphable Counters \cite{saileshwar2018morphable} pack 64-128 VNs per block to improve efficiency.

\begin{figure}[t]
  \centering
  \includegraphics[scale=0.48]{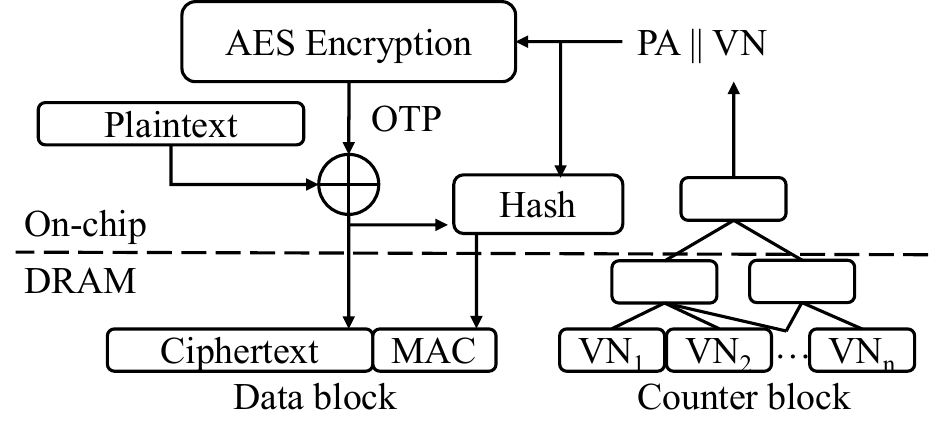} 
  \caption{SGXv1-like memory protection scheme}
  \label{fig:Counter_Mode_Encryption}
  \vspace*{-5mm}
\end{figure}

The encryption process, shown in Fig. \ref{fig:Counter_Mode_Encryption}, creates a one-time pad (OTP) by encrypting the data's physical address (PA) concatenated with its VN:
\[ Ciphertext = Plaintext \oplus \text{AES\_Enc}(PA \Vert VN)\]
This OTP is XORed with plaintext to produce ciphertext, providing confidentiality. Intel and AMD have integrated specialized hardware for this operation, enabling transparent memory encryption widely used by cloud providers \cite{intel_white_paper,amd_white_paper,encrypted_NVM_hpca_2018}.

For integrity protection, a MAC verifies data authenticity:
\[MAC = Hash (Ciphertext \Vert (PA \Vert VN))\]
During reads, the system recomputes and verifies the MAC. However, MAC verification alone cannot prevent replay attacks where adversaries reinsert old data with its matching VN and MAC. To address this, MT \cite{Merkletree} create a hierarchical integrity structure with VNs as leaves and intermediate hash values, culminating in an on-chip root hash. Any tampering with VNs is detected during tree traversal, ensuring comprehensive protection against replay attacks.

\subsection{Secret Sharing}
\label{bg:ss}
Secret sharing is a cryptographic technique that divides data into multiple shares, offering an alternative to traditional encryption for memory protection \cite{SS_survey}. In Shamir's Secret Sharing (SSS), the most widely used scheme, a secret $s$ is encoded into a polynomial $f(x) = s + a_1x + a_2x^2 + \ldots + a_{k-1}x^{k-1}$ with randomly chosen coefficients $a_i$ \cite{SSS,SSS1994}. Shares are generated by evaluating this polynomial at distinct points, and the original secret can be reconstructed from any $k$ shares using polynomial interpolation such as Lagrange or Barycentric methods \cite{interpolation}. The $(k, n)$ scheme ensures perfect secrecy—splitting a secret into $n$ shares where fewer than $k$ shares reveal no information about the original data \cite{SSS}. This property has made secret sharing valuable in diverse applications from quantum communications to blockchain systems \cite{SS_quantum, SS_quantum2, SS_block_chain, SS_block_chain2}, demonstrating its robustness for preserving data privacy in memory systems \cite{SS_block_chain}.

%% file: 03_Motivation.tex
\section{Observations and Motivations}
\label{sec:Motivation}


\HPCA{In this section, we analyze the security and performance of existing memory encryption schemes, highlight their key limitations, review current solutions and their drawbacks, and present our motivation for addressing these challenges.}

\subsection{Analysis on Secure Memory}

AES-XTS is widely adopted in platforms such as Intel SGX v2, Intel TDX, and AMD SEV due to its simplicity and efficiency. However, AES-XTS suffers from major security limitations, including a lack of prevention of replay attacks and integrity verification \cite{XTS_sca1,XTS_sca2}. These vulnerabilities make it less suitable for environments that require strong resistance to manipulation and data freshness guarantees.

AES-GCM improves security by providing built-in authentication and integrity verification through its authenticated encryption design. Although it prevents many tampering and substitution attacks, AES-GCM still lacks robust replay protection and introduces practical challenges such as nonce management, sequential processing requirements, and tag computation overhead \cite{GCM_WEAK1,GCM_WEAK2}. These factors can complicate deployment and limit scalability, especially in large memory systems.

In contrast, SGXv1-like offers the most comprehensive security, protecting against a broad range of attacks including replay attacks. This scheme was adopted in earlier secure memory implementations like Intel SGX v1. However, these strong guarantees come at the cost of significant performance bottlenecks, such as high latency due to MT traversal, limited secure memory size (e.g. 128MB), and poor scalability as memory capacity grows. These limitations have led Intel to abandon SGXv1-like in subsequent generations of SGX, mainly due to its impact on performance and the difficulty of supporting larger memory regions~\cite{sgx_v2}. As a result, SGXv1-like has become impractical for protecting large memory in performance-sensitive applications. In the next section, we analyze the performance bottlenecks of SGXv1-like using experimental evaluation.

\begin{figure}
\centering
  \vspace*{-5mm}
 \resizebox{0.95\columnwidth}{!}{\includegraphics{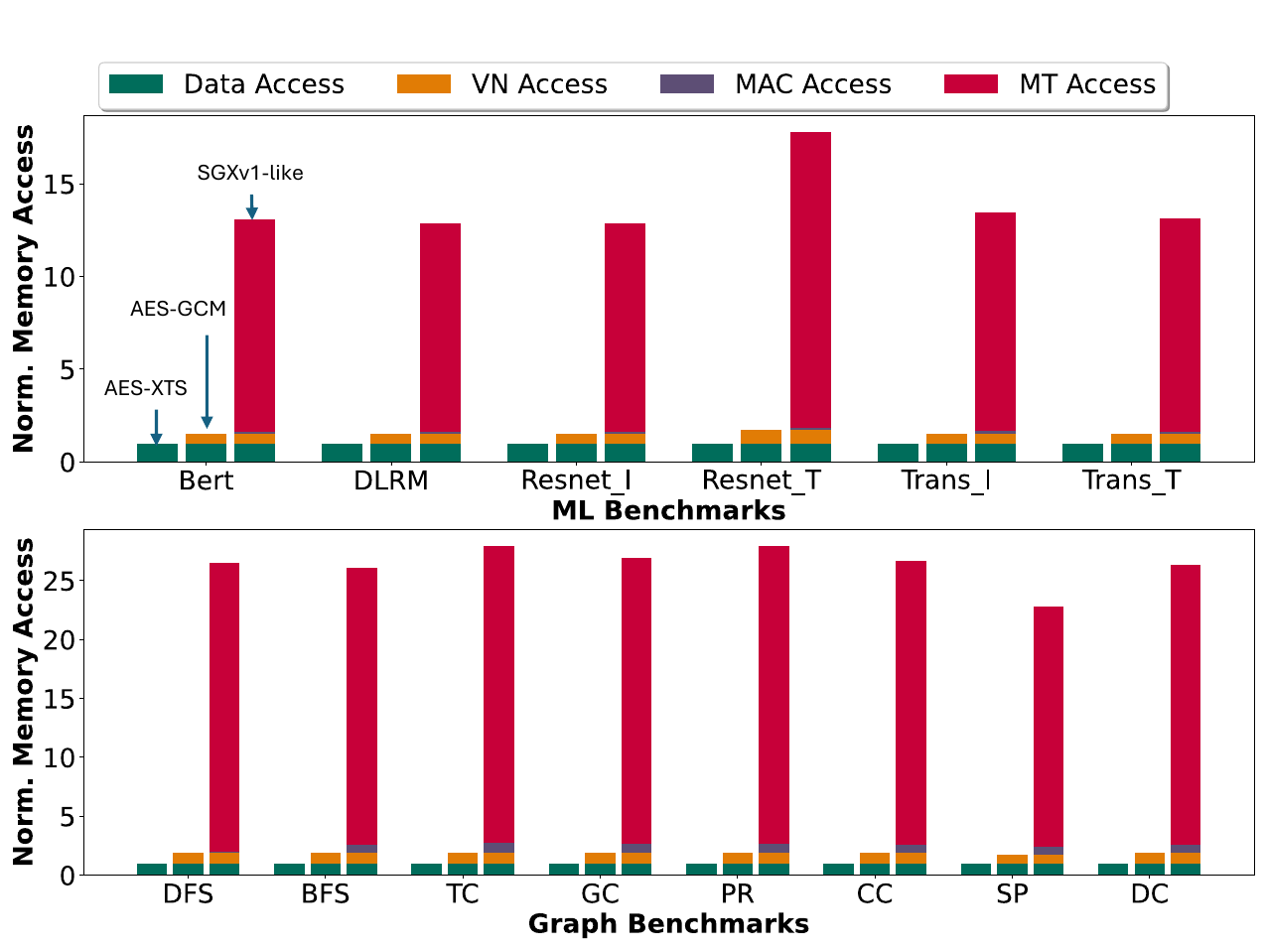}}
 \vspace*{-4mm}
\caption{Normalized memory access overhead of encryption schemes. (a) ML Benchmarks and (b) Graph Benchmarks. AES-XTS is the left bar, AES-GCM is the middle bar, and SGXv1-like is the right bar. All values are normalized to data access.}
\label{fig:Memory_analysis}
\vspace*{-6mm}
\end{figure}

\subsubsection{Secure Memory Performance Analysis}

\HPCA{To better understand the performance of each secure memory encryption scheme, we perform detailed memory access analysis by simulating different secure memory systems. For AES-XTS, we simulate the Intel SGX v2 setup \cite{sgx_v2}; for AES-GCM, we base our simulation on CXL.mem specifications \cite{cxl_1}; and for SGXv1-like, we model the system according to Intel SGX v1~\cite{SGX}. All simulations are conducted using Gem5~\cite{gem5}. The system is configured as a 4-core setup with an 8MB shared last-level cache (LLC), a 128KB VN cache, and with a relatively large 32 GB secure memory size (see Sec~\ref{sec:Evaluation} for details).}

To evaluate performance under diverse memory access behaviors, we use two benchmark categories: machine learning (ML) workloads with regular access patterns~\cite{ML_seq,ML_seq2} and graph applications with irregular patterns~\cite{graph_ran1,graph_ran2}. 
For graph workloads, we employ GraphBIG~\cite{graphbig} with the GitHub social network dataset~\cite{graph_data_set}, evaluating depth-first search (DFS), breadth-first search (BFS), triangle counting (TC), graph coloring (GC), page ranking (PR), connected components (CC), shortest path (SP), and degree centrality (DC) using four threads. 
For ML workloads, we include BERT~\cite{BERT} (sequence length 128, hidden size 768), DLRM~\cite{DLRM} (100K embeddings, 16-dim), ResNet50 inference/training (ResNet\_I/T, input: 64$\times$56$\times$56, kernel: 3$\times$3), and Transformer inference/training (Trans\_I/T, sequence: 128, embedding: 512, 8 heads). 
Each benchmark runs a single inference or one training epoch in gem5 to ensure feasible simulation time.

Fig.~\ref{fig:Memory_analysis} shows the normalized memory access overhead across encryption schemes, with all values normalized to data accesses (set to 1). AES-XTS introduces minimal overhead since it directly uses memory addresses as tweaks without requiring additional DRAM metadata, while AES-GCM adds moderate overhead from version-number (VN) accesses. In contrast, SGXv1-like designs incur substantial overhead due to integrity verification via Merkle Tree (MT) traversal, particularly when protecting large memory capacities. For example, securing 32~GB of memory using Morphable Counters~\cite{saileshwar2018morphable} requires one VN per 128 cache lines, yielding about $(32,\text{GB})/(64,\text{B}\times128)\approx4\text{M}$ VNs and roughly $\log_{8}(4\text{M})\approx7$ tree levels for an 8-ary MT~\cite{saileshwar2018morphable,horus}. Each VN verification thus fetches multiple sibling nodes across these levels, resulting in an average of $\sim$20 additional memory accesses per verification~\cite{saileshwar2018morphable,synergy,toleo,softvn}. Although regular workloads such as ML benefit from caching locality, irregular memory patterns in graph and database applications often trigger worst-case MT traffic. \emph{This memory access amplification translates to roughly 30\% end-to-end performance degradation compared to unprotected execution, consistent with observations in prior works including EMCC \cite{emcc}, RMCC \cite{rmcc}, MGX \cite{mgx}, SoftVN \cite{softvn}, CounterLight \cite{counterlight}, and COSMOS \cite{cosmos}.}


\HPCA{Based on our analysis and experimental results, \textbf{our first motivation is to achieve the security guarantees of SGXv1-like, maintaining strong data integrity and replay protection, while reducing its memory access overhead.}}

\subsection{Current Solutions and Their Limitations}

\HPCA{Numerous research efforts have targeted the performance bottleneck of SGXv1-like in secure memory systems, primarily addressing the overheads from VNs. These approaches can be categorized into four distinct strategies:}

\HPCA{\textbf{(1) Algorithm-Level Optimizations:} Techniques such as Morphable Counters (MorphCTR) \cite{saileshwar2018morphable} and Synergy \cite{synergy} improve VN caching efficiency by encoding more VNs per memory block or colocating MACs with data. However, despite these improvements, they still incur substantial overhead from frequent integrity verification due to MT traversals on cache misses.}

\textbf{(2) Architectural Enhancements:} 
Approaches such as EMCC~\cite{emcc}, RMCC~\cite{rmcc}, and COSMOS~\cite{cosmos} mitigate VN cache miss latency by introducing on-chip optimizations—either caching VNs in lower-latency levels (e.g., L2) or leveraging prediction and reuse mechanisms to anticipate future VN accesses. While these methods effectively reduce VN miss penalties, they introduce additional hardware complexity, timing closure challenges, and verification overhead due to deeper integration with core pipeline and cache hierarchies.

\HPCA{\textbf{(3) Eliminating Off-chip VN Storage:} Solutions such as Osiris \cite{osiris}, MGX \cite{mgx} and SoftVN \cite{softvn} avoid storing VNs off-chip entirely, leveraging internal accelerator states or software-managed loop counters for deriving VN values. Although this fully eliminates integrity verification overhead, the reliance on potentially compromised software or hardware states introduces critical security vulnerabilities and scalability limitations.}

\HPCA{\textbf{(4) Hybrid Counter-Based:} Methods like Counter-Light (CTR-L) \cite{counterlight} combine SGXv1-like for infrequently accessed data and AES-XTS for frequently accessed data to balance security and performance. While effective in reducing overheads, the data blocks encrypted using AES-XTS remain susceptible to replay attacks, weakening overall security guarantees.}

\subsection{Rethinking Secure Memory Design }

\begin{figure*}[t]
	\centering
	\includegraphics[width=0.93\textwidth]{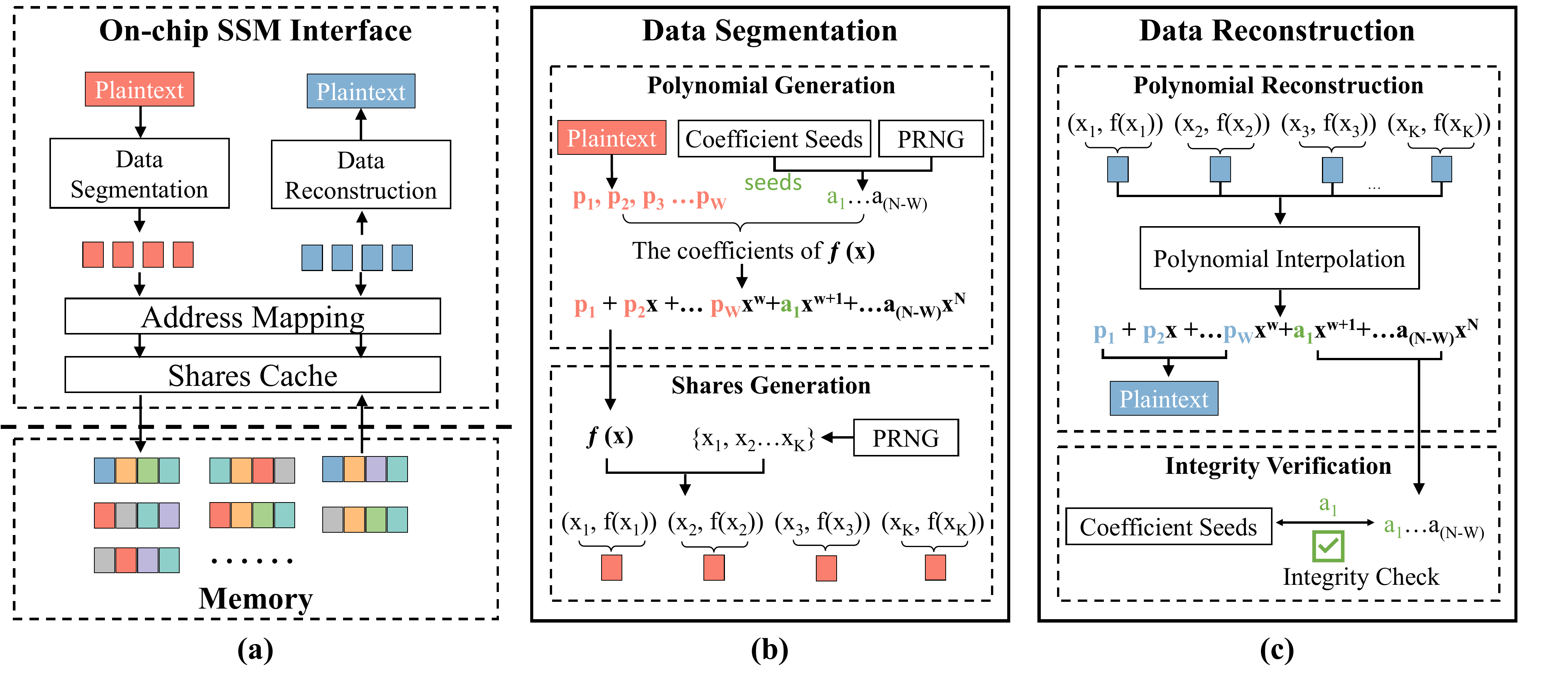}
   \vspace*{-3mm}
	\caption{The high-level overview of (a) SSM design, (b) data segmentation process, (c) data reconstruction process.}
    \label{fig:ssm_OVERALL}
     \vspace*{-5mm}
    
\end{figure*}

\HPCA{Current solutions to SGXv1-like's performance bottlenecks all operate within the constraints of traditional encryption, accepting the fundamental separation between data and security metadata (VNs, MACs, and MT nodes). We challenge this assumption: \textbf{Why must security metadata exist separately from the data it protects?} If transformed data could inherently provide all necessary security properties, we could eliminate the overhead of managing separate metadata structures entirely.}

\HPCA{This insight leads us to secret sharing \cite{additive_secrate_sharing,SS_survey,SSS1994}, where data are divided into shares that inherently provide multiple security properties. Individual shares reveal nothing about the original data, yet any tampering causes reconstruction to fail, providing both confidentiality and integrity without separate metadata.}

\HPCA{By applying secret sharing to memory protection, we can eliminate the overhead of managing VNs, MACs, and MT nodes entirely. The shares themselves serve as both encrypted data and an integrity verification mechanism. \textbf{Our second motivation is to design a secret-sharing-based memory encryption scheme that achieves the security guarantees of SGXv1-like while eliminating its metadata management overhead.}}

%% file: 04_Secure_Scattered_Memory.tex
\section{Secure Scattered Memory Architecture}
\label{sec:Secure Scattered Memory}

\HPCA{Building on our key motivations from Section \ref{sec:Motivation}, we propose SSM.} SSM is an innovative approach to securing data storage. Rather than storing complete data, SSM divides information into secret shares and distributes them across the memory. This strategy ensures that memory content consists solely of shares, making individual memory lines hold data fragments that are meaningless without the entire set. 

\begin{table}[t]
\centering
\scriptsize
\caption{Comparison of Security Boundaries and Protection Guarantees among Secure-Memory Designs.
(C: Confidentiality, I: Integrity, F: Freshness, A: Access-Pattern Protection)}
\label{tab:threat_model_comparison}
\renewcommand{\arraystretch}{1.1}
\setlength{\tabcolsep}{4pt}
\begin{tabular}{lcccccc}
\toprule
\textbf{Scheme} & \textbf{Security Boundary} & \textbf{C} & \textbf{I} & \textbf{F} & \textbf{A} \\ 
\midrule
\textbf{SGXv1}~\cite{SGX_explain} & \makecell[c]{On-chip trusted (cores, caches, MC);\\ Off-chip memory  \& OS untrusted} & \cmark & \cmark & \cmark & \xmark$^{\P}$  \\
\textbf{SGXv2}~\cite{sgx_v2} & Same as SGXv1 & \cmark & \xmark & \xmark & \xmark  \\

\textbf{MorphCTR}~\cite{saileshwar2018morphable} &
Same as SGXv1 &
\cmark & \cmark & \cmark & \xmark \\
\textbf{EMCC}~\cite{emcc} & Same as SGXv1  & \cmark & \cmark & \cmark & \xmark \\
\textbf{RMCC}~\cite{rmcc} & Same as SGXv1  & \cmark & \cmark & \cmark & \xmark  \\
\textbf{COSMOS}~\cite{cosmos} & Same as SGXv1  & \cmark & \cmark & \cmark & \xmark  \\
\textbf{Counter-Light}~\cite{counterlight} & Same as SGXv1  & \cmark & \cmark & \textcolor{orange}{\textbf{$\circ$}}$^{\ddagger}$ & \xmark \\
\textbf{SoftVN}~\cite{softvn} &\makecell[c]{SGXv1-based; SoftVN assumes a \\trusted OS and enclave software$^{*}$} & \cmark & \cmark & \cmark & \xmark  \\
\textbf{SSM (This work)} & Same as SGXv1  & \cmark & \cmark & \cmark & \textcolor{orange}{\textbf{$\circ$}}$^{\dagger}$  \\
\bottomrule
\end{tabular}
\vspace{1mm}
\begin{flushleft}
\scriptsize{
$^{\P}$~Access-pattern leakage is considered out of scope in SGXv1 \\
$^{\ddagger}$~Freshness partially ensured; AES-XTS mode weakens replay protection for hot data. \\ 
$^{*}$~Extends the SGXv1 model by delegating VNs management to enclave software. \\
$^{\dagger}$~Access pattern observable but statistically unexploitable. }
\end{flushleft}
\vspace{-5mm}
\end{table}

\subsection{Threat Model} \label{sec:SSM_Threat_Model}
SSM adopts the widely used SGXv1-based threat model followed by counter-based secure-memory systems~\cite{sok,sgxaxe, sgx_security, saileshwar2018morphable, emcc, rmcc, counterlight}.
As summarized in Table~\ref{tab:threat_model_comparison}, the trusted components include all on-chip components (CPU cores, caches, and the memory controller). 
On the other hand, the off-chip memory, memory bus, and system software are untrusted. An adversary with full control of the operating system (OS) can arbitrarily read or modify memory contents and replay previously observed ciphertexts. 
However, the adversary cannot compromise on-chip hardware or extract on-chip secrets.

Under this model, SSM guarantees the standard secure-memory properties: 
\emph{confidentiality}, \emph{integrity}, and \emph{freshness} of all off-chip data. Confidentiality ensures that the attacker cannot infer plaintext information from encrypted data. Integrity prevents unauthorized modification of data blocks. Freshness ensures that stale or replayed data cannot be accepted as valid. In contrast, \emph{access-pattern protection} is considered out of scope, following the same assumption as all listed schemes in Table~\ref{tab:threat_model_comparison}. Nevertheless, SSM introduces randomized share placement that statistically decorrelates access sequences from logical data usage, making observable patterns statistically unexploitable. The detailed rationale and security analysis explaining why SSM achieves this property and ensures overall system security are presented in Section~\ref{sec:Security_analysis}.

\begin{table}
\centering
\caption{List of parameters used in SSM.}
\begin{tabular}{c|l}
\hline
\textbf{Symbol} & \textbf{Definition} \\
\hline
$N$ & Polynomial degree \\
\hline
$K$ & Total numbers of shares read/written per memory access\\
\hline

$t$ & Minimal number of shares\\ & required to reconstruct the original data \\
\hline
$d$ & The number of unused shares \\& in each memory access \\
\hline
\end{tabular}
\label{tab:SSM_symbols}
 \vspace*{-5mm}
\end{table}




\subsection{SSM Overview}
Fig. \ref{fig:ssm_OVERALL}(a) depicts a high-level overview of SSM and Table \ref{tab:SSM_symbols} lists key parameters that will be frequently referenced in subsequent sections.
The SSM system consists of the shares cache, data segmentation, and data reconstruction, integrated within the memory controller.

\textbf{Data segmentation/reconstruction modules:} These modules handle breaking down data blocks into shares during write operations and reconstructing the original data from the shares during read operations.
Upon each memory access, SSM first checks the shares cache for the required shares. 
If the shares cache does not contain all the needed shares, SSM reads or writes a total of $K$ shares from or to the memory. Out of these $K$ shares, $t$ shares are necessary to reconstruct the original data, while the remaining $d$ shares are unused shares used to obfuscate access patterns and prevent attackers from reconstructing the data. The unused shares are then stored in the shares cache.

\textbf{Shares Cache:} Each memory access in SSM involves multiple memory blocks, but only a portion of the shares are used to reconstruct the required data block. 
The remaining shares, acting as unused values to deter attackers, are valid shares from other data blocks. 
SSM employs a shares cache (cache for storing secret shares) to store fetched data, enhancing both performance and security. Each memory request (read or write-back in case of dirty write) in SSM is separated into $K$ share requests to DRAM, all of which go to the shares cache first. If the shares cache hits, it uses the share from the cache. If there is a miss in the shares cache, it accesses the DRAM for the corresponding share.


To maximize the effectiveness of the shares cache, data blocks with contiguous addresses are mapped into the same memory blocks. This strategy leverages the locality of consecutive data blocks, which likely exhibit similar access patterns \cite{denning2005locality}. 
For example, if data blocks $d_1$ and $d_2$ are consecutive, they are segmented into shares ($d_1 \rightarrow sa_1, sa_2, sa_3, \ldots$ and $d_2 \rightarrow sb_1, sb_2, sb_3, \ldots$, where $sa_x$ and $sb_x$ represent the x-th share of $d_1$ and $d_2$, respectively.) and map shares of $sa_x$ and $sb_y$ (of arbitrary x and y value) into the same memory block, i.e.,$s_1 \rightarrow sa_1, sb_2, \ldots$; $s_2 \rightarrow sa_2, sb_3, \ldots$; $s_3 \rightarrow sa_3, sb_1, \ldots$). This method prefetches data blocks with similar locality during SSM accesses, improving the shares cache hit rate. 

Implemented as a physically addressed cache with an LRU replacement policy, the shares cache is accessed after LLC misses and address translation. Given $K$ (listed in Table \ref{tab:SSM_symbols}), $K$ consecutive data blocks are mapped to $K$ share blocks in memory, leveraging the principle that consecutive blocks typically exhibit similar access patterns. The impact of $K$ and the shares cache hit rate is explored in §\ref{sec:Experimental Results}.

\subsection{Data Segmentation}
\label{sec: ds}

Data segmentation in SSM aims to: (i) segment data into shares such that individual shares reveal no meaningful information about the original data, and (ii) ensure the segmentation is reversible for accurate data reconstruction when authorized. 
This process leverages the principles of SSS \cite{SSS} of using polynomials for data segmentation, fulfilling these two goals. 
Fig. \ref{fig:ssm_OVERALL}(b) shows the data segmentation process, which consists of two phases: (1) polynomial generation and (2) shares generation.
\subsubsection{\textbf{Polynomial Generation}}
A polynomial of degree \( N \) is formulated. Initially, the original data block is divided into segments \( p_{1} \) to \( p_{W} \). \HPCA{To enhance reconstruction accuracy, each segment is padded with additional zero bits before forming the polynomial.} The  polynomial is represented as:
\[ p_{1} + p_{2} \times x + \dots + p_{W} \times x^W + a_{1} \times x^{W+1} \dots + a_{N-W} \times x^N \]
where the segments \( p_{1} \) to \( p_{W} \) act as the coefficients from degree 0 to degree \( W \). A set of coefficients \( a_1, a_2, \dots, a_{N-W} \) is derived from securely stored coefficient seeds to facilitate integrity verification.  

\subsubsection{\textbf{Shares Generation}} Random values \( x_1 \) to \( x_K \) are produced using A pseudo-random number generator (PRNG). The polynomial is evaluated for each value to determine its \( f(x) \). 
Each resulting pair, e.g. \((x_1, f(x_1))\), is referred to as a share. Using this method, a total of \( t \) shares are generated for the polynomial of each block and subsequently stored in memory.


\subsection{Data Reconstruction}
\label{sec:Data_reconstruction}

Data reconstruction in SSM aims to: (i) accurately recover the original data from a subset of shares, and (ii) verify the integrity of the reconstructed data. This process leverages the polynomial-based structure established during data segmentation to achieve both goals. 
Fig. \ref{fig:ssm_OVERALL}(c) illustrates the data reconstruction process. The process encompasses two main phases: (1) polynomial reconstruction and (2) integrity verification.

\subsubsection{\textbf{Polynomial Reconstruction}}
During each memory read operation in SSM, $K$ shares are retrieved, employing $t$ shares for polynomial reconstruction and $d$ shares to obscure the access pattern ($d > t$). To reconstruct the original polynomial, a minimum of $t$ shares is necessary. 

\HPCA{SSM performs all arithmetic operations in the Galois Field, ensuring error-free encryption and decryption. In this finite field, addition is XOR and multiplication follows polynomial arithmetic modulo an irreducible polynomial, enabling well-defined division operations. We employ Lagrange interpolation to recover the polynomial coefficients (original data) from the shares $\{(x_1, f(x_1)), (x_2, f(x_2)), \ldots, (x_t, f(x_t))\}$:}

\begin{equation}
    f(x) = \sum_{i=1}^{t} f(x_i) \prod_{j=1, j \neq i}^{t} \frac{x - x_j}{x_i - x_j}
\end{equation}

\HPCA{The polynomial coefficients $p_1, p_2, \ldots, p_W$ directly form the original data segments, with each coefficient representing one byte of the 64-byte data block. After reconstruction, the padding is removed from each segment to recover the original 64-byte data. In Section~\ref{sec:Experimental Results}, we experimentally evaluate the accuracy of data reconstruction using Galois field computations.}

\subsection{Data Segmentation and Reconstruction Example} 
\label{sec:SSM_example}
To help understand the scheme, we give an example of the data segmentation and reconstruction here: \textbf{Segmentation:} \circled{1} Split 64-byte block into eight 8-byte segments $p_1$-$p_8$. \circled{2} Pad each with 8 zero bytes. \circled{3}Form polynomial $f(x) = p_1 + p_2x + \ldots + p_8x^8 + a_1x^9$ where $a_1$ is from coefficient seed. \circled{4} Generate shares: evaluate $f(x)$ at random $x_i$. \circled{5} Store as 9-byte $(x_i, f(x_i))$ pairs.

\textbf{Reconstruction:} \circled{1} Retrieve 10 shares. \circled{2} Use Lagrange interpolation in recovery of $f(x)$. \circled{3} Extract $p_1$-$p_8$; verify seed $a_1$. \circled{4} Remove padding. \circled{5} Concatenate to restore 64 bytes. Memory stores 7 shares/block; $8$ blocks are fetched during each access, providing a total of $56$ shares (K=56) per read/write.




\subsection{Key Design Considerations in SSM}
\label{sec:keydesign}
We elaborate below on key aspects of our design that make SSM a robust, secure, and efficient memory system.

\begin{figure}[tb]
\centering
\scriptsize 
\begin{tikzpicture}[
    box/.style={draw, rectangle, minimum height=0.5cm, minimum width=1.3cm, font=\scriptsize, align=center},
    tlb/.style={box, fill=red!10},
    arrow/.style={->, >=stealth}
]
\node[box] (addr) {Physical\\Address};
\node[tlb, right=0.5cm of addr] (tlb) {SSM\\TLB};
\node[box, right=0.5cm of tlb] (walk) {SSM Page\\Walk};
\node[box, right=0.5cm of walk] (blocks) {K Block\\Addresses};

\draw[arrow] (addr) -- (tlb);
\draw[arrow] (tlb) -- (walk);
\draw[arrow] (walk) -- (blocks);
\end{tikzpicture}

\vspace{0.15cm}
\scriptsize Consecutive addresses (A, B, C, D) map to same K blocks:

\vspace{0.1cm}
\begin{tikzpicture}[
    block/.style={draw, rectangle, minimum height=0.6cm, minimum width=1.2cm, font=\scriptsize, align=center}
]
\node[block] at (0,0) {Block 0\\{\tiny [s$_0$,s$_1$,...,s$_n$]}};
\node[block] at (1.5,0) {Block 1\\{\tiny [s$_0$,s$_1$,...,s$_n$]}};
\node[block] at (3,0) {Block 2\\{\tiny [s$_0$,s$_1$,...,s$_n$]}};
\node at (4.2,0) {\scriptsize ...};
\node[block] at (5.5,0) {Block K-1\\{\tiny [s$_0$,s$_1$,...,s$_n$]}};
\end{tikzpicture}

\vspace{0.1cm}
\centerline{\scriptsize \textit{offset} = (addr / line\_size) \% n; \quad \textit{share}[i] = (offset + i) \% n}

\vspace{0.1cm}
\scriptsize
\setlength{\tabcolsep}{3pt} 
\begin{tabular}{|c|c|c|c|c|c|c|}
\hline
\textbf{Addr} & \textbf{Off} & \textbf{Blk 0} & \textbf{Blk 1} & \textbf{Blk 2} & \textbf{...} & \textbf{Blk K-1} \\
\hline
A & 0 & \cellcolor{yellow!30}s$_0$ & \cellcolor{yellow!30}s$_1$ & \cellcolor{yellow!30}s$_2$ & ... & \cellcolor{yellow!30}s$_{K-1}$ \\
B & 1 & \cellcolor{yellow!30}s$_1$ & \cellcolor{yellow!30}s$_2$ & \cellcolor{yellow!30}s$_3$ & ... & \cellcolor{yellow!30}s$_0$ \\
C & 2 & \cellcolor{yellow!30}s$_2$ & \cellcolor{yellow!30}s$_3$ & \cellcolor{yellow!30}s$_4$ & ... & \cellcolor{yellow!30}s$_1$ \\
D & 3 & \cellcolor{yellow!30}s$_3$ & \cellcolor{yellow!30}s$_4$ & \cellcolor{yellow!30}s$_5$ & ... & \cellcolor{yellow!30}s$_2$ \\
\hline
\end{tabular}
\caption{SSM address mapping.}
\label{fig:ssm-address-mapping}
\vspace*{-5mm}
\end{figure}

\subsubsection{Feasibility} 
\label{sec:address-mapping}
\textbf{Address Mapping:} SSM presents a unique challenge in address mapping: it requires managing a one-to-many relationship between virtual and physical addresses, as each original data block is split into multiple share blocks in DRAM. We address this challenge through a two-level mapping scheme illustrated in Fig.~\ref{fig:ssm-address-mapping}.

Upon an LLC miss, SSM intercepts the physical address and performs an SSM-specific page walk that returns K physical addresses, one for each required share block. A key innovation is that consecutive data addresses map to the same K physical blocks to exploit spatial locality.

The critical insight is that while security requires distributing shares across memory, performance demands cache efficiency. We resolve this tension through a rotating share selection pattern. As shown in Fig.~\ref{fig:ssm-address-mapping}, addresses A, B, C, and D all map to the same K blocks, but each selects different shares using the formula: \texttt{share\_index[i] = (address\_offset + i) \% n}, where n is shares per block.

\textbf{Mapping Metadata Protection:}
Protection applies to enclave data: EPC pages are encrypted and integrity-checked by the CPU (MEE/EPCM) when off-chip~\cite{Valut}. 
SSM follows this model: mapping metadata in DRAM is unprotected, while all off-chip share blocks are encrypted and verified by the on-chip controller.



\textbf{Metadata Management:} \HPCA{All SSM operations—data segmentation, reconstruction, and address offset computation—execute within the MC after LLC misses. Coefficient seeds are stored on-chip in the MC, while SSM page tables reside in DRAM.}

\subsubsection{Security}
\label{sec:security_IV}
\textbf{Integrity Verification:} Integrity verification utilizes coefficient seeds that are pre-stored, and periodically updated on-chip, balancing security needs with limited storage capacity. Multiple data blocks can share a single seed, significantly reducing on-chip storage requirements. During data reconstruction, specific polynomial coefficients (\( a_1 \) to \( a_{N-W} \)) are checked against on-chip stored coefficient seeds to ensure that each coefficient (\( a_1 \) to \( a_{N-W} \)) matches its corresponding pre-determined seed value. As shown in Fig.~\ref{fig:ssm_OVERALL}(b) and (c), the coefficient(\( a_1 \) to \( a_{N-W} \)) are verified against the on-chip seed. Integrity checking occurs for every reconstruction. Matching coefficients and seeds indicate integrity, while deviations suggest potential data tampering.

\textbf{Replay Attack Prevention:} \HPCA{SSM prevents replay attacks through dynamic share relocation, eliminating the need for version numbers or MT. As illustrated in Fig.~\ref{fig:replay-protection}, every write operation remaps shares to new physical locations and updates the SSM page table accordingly. This creates an effective barrier against replay attacks: Even if an attacker captures valid shares at time T1 from locations [1,2,3,4], these become useless after the system relocates the shares to [5,6,7,8] at time T2. The attacker faces two insurmountable challenges: they cannot determine the new locations without accessing the secure SSM page table, and any attempt to inject old shares into the old locations fails because SSM reads from the updated locations. This location-based defense transforms the page table into an implicit freshness indicator: only the current table knows where valid shares reside.}

\begin{figure}[t]
\centering
\footnotesize
\setlength{\tabcolsep}{3pt}
\renewcommand{\arraystretch}{1.2}
\begin{tabular}{@{}l@{}}
\textbf{SSM Replay Attack Prevention via Dynamic Relocation} \\[4pt]
\begin{tabular}{@{}ll@{}}
T1: & Initial write: Shares stored at Loc[1,2,3,4] \\ 
T2: & Update: \sout{Loc[1,2,3,4]} $\rightarrow$ \colorbox{green!20}{Loc[5,6,7,8]} + Page Table Update \\ 
T3: & Attacker replays old shares $\rightarrow$ Old Loc[1,2,3,4] \\ 
T4: & SSM reads from New Loc[5,6,7,8] (per updated page table) \\ 
\end{tabular}\\[2pt]
\textbf{Result:} Attack fails—old shares at wrong locations
\end{tabular}
\caption{SSM prevents replay attacks through dynamic share relocation.}
\label{fig:replay-protection}
\vspace*{-5mm}
\end{figure}

\subsubsection{Efficiency} \HPCA{SSM addresses its two primary performance bottlenecks—additional page table walks and increased share accesses—through strategic caching and intelligent address mapping.}

\HPCA{To mitigate memory access overhead, SSM employs two specialized caches. The SSM TLB caches page table mappings that translate data addresses to share locations, reducing repeated page walks. The shares cache stores recently accessed shares within the memory controller, serving subsequent requests without DRAM access.}

\HPCA{The key efficiency innovation lies in mapping consecutive addresses to the same K physical blocks while using different share indices, as described in Section~\ref{sec:address-mapping}. This design exploits spatial locality: When accessing address B after A, the SSM TLB already contains the required mappings, while the shares cache likely holds shares from the same physical blocks. Our experiments demonstrate the effectiveness of this approach, with consecutive mapping achieving significantly higher cache hit rates compared to random mapping. This synergy between intelligent mapping and caching enables SSM to approach traditional memory system performance.}

%% file: 05_Threat_Model.tex
\section{Security Analysis}
\label{sec:Security_analysis}
This section analyzes how SSM satisfies the confidentiality, integrity, and freshness guarantees defined in the SGXv1-based threat model (§\ref{sec:SSM_Threat_Model}), and demonstrates that SSM securely resists data reconstruction and achieves statistically unexploitable access patterns, without violating any theoretical assumptions of secret sharing.

\subsection{Security Guarantees Analysis}
\subsubsection{Confidentiality Guarantee Analysis}
SSM achieves information-theoretic confidentiality through its $t$-out-of-$N$ secret-sharing-based encoding. 
Each data block is mathematically embedded into the coefficients of a function $f(x)$ over a finite field, and the memory stores its corresponding $(x, f(x))$ shares.
Any subset of fewer than $t$ shares reveals no information about the underlying data, as all partial combinations are statistically indistinguishable from random values. 
The relationship among shares is determined by an on-chip mapping that never leaves the trusted boundary, making it infeasible for an adversary to correlate or randomly select the correct $t$ shares from the entire memory space. Consequently, plaintext recovery without on-chip information is statistically impossible.

\subsubsection{Integrity Verification Analysis}
SSM ensures data integrity through coefficient verification inherently embedded in the polynomial reconstruction process~(\S\ref{sec:security_IV}). Each reconstruction must yield a function whose coefficients precisely match the authenticated on-chip seeds. This design elegantly couples correctness and integrity: if and only if all $t$ shares are genuine, the polynomial can be solved to produce consistent coefficients. Any tampering of off-chip shares leads to mismatch during verification, as seeds are unpredictable and inaccessible to the adversary. The chance of a forged block passing verification is negligible, providing MAC-equivalent integrity without maintaining per-line counters or tags.

\subsubsection{Freshness and Replay Resistance Analysis}
Replay attacks attempt to reintroduce stale data into memory so that the system consumes outdated values. SSM guarantees freshness by regenerating shares and remapping them to new physical locations on each write. Conceptually, the regenerated shares in SSM serve a role analogous to the incremented counters in SGXv1: both represent a new, authenticated version of data. Even if an adversary captures valid shares at time $T_1$, these become useless after remapping at $T_2$, as both their locations and encoding have changed. Any replay of old shares leads to reconstruction failure due to mismatched mappings.

\subsection{Data Reconstruction Analysis}
While most secure-memory systems consider access-pattern protection out of scope~(Table~\ref{tab:threat_model_comparison}), SSM revisits this issue from its secret-sharing foundation.
In theory, the confidentiality of secret sharing relies on the non-collusion assumption, which assumes at least one honest participant and does not disclose its share, preventing an adversary from collecting $t$ valid shares. 
SSM realizes the same principle architecturally: all shares are generated through a trusted on-chip SSM interface and are stored off-chip as independent, information-theoretically indistinguishable values. Each share carries no identity linking it to others, and the relationships among them are determined solely by the on-chip mapping, which remains inaccessible to the adversary. Hence, all off-chip shares are \emph{honest by construction}, providing the system-level equivalent of non-collusion.

Beyond this theoretical equivalence, SSM reinforces reconstruction resistance through dynamic regeneration of shares and periodic remapping of their physical locations.  The coefficient seeds act as the trusted anchor, ensuring that only authentic shares corresponding to the current mapping can pass verification. Even if all off-chip shares are exposed to an attacker, the coefficient seeds—securely stored and never exported from the on-chip SSM module—remain unknown, preventing any attempt to reconstruct the original data.

Because regeneration preserves coefficients but renews share locations, the physical distribution of shares constantly changes, eliminating temporal stability and preventing linkage across memory access patterns. Even if an adversary observes access traces, the randomized placement and remapping ensure that no single share can be persistently tracked, nor can multiple shares of the same \( f(x) \) be correlated. Consequently, both data reconstruction and access-pattern exploitation become statistically infeasible, maintaining the theoretical confidentiality of secret sharing within a single-system architecture. 

In SSM, each memory access touches $K$ physical shares, of which exactly $t$ are genuine and $K{-}t$ are decoys. An adversary observing the $K$ accessed shares but lacking the on-chip mapping must guess which $t$ are genuine. The per-access success probability is
\begin{equation}
P_{\text{access}} = \frac{1}{\binom{K}{t}}
\end{equation}
For $K{=}56$ and $t{=}10$, this yields $P_{\text{access}} \approx 2.8\times 10^{-11}$. 
Across $M$ independent accesses (e.g., writes with regeneration/remapping), the cumulative success is
\begin{equation}
P_{\text{total}} = 1 - (1 - P_{\text{access}})^M \approx M\,P_{\text{access}} \quad (P_{\text{access}}\approx 0)
\end{equation}

Even though access patterns in SSM are observable, they remain statistically unexploitable, making the probability of successful data reconstruction negligible.

%% file: 07_Methology.tex
\section{Implementation and Evaluation Methodology}
\label{sec:Evaluation}

\subsection{Hardware Component Implementation}

\begin{figure} [t] 
  \centering
 \resizebox{1.00\columnwidth}{!}{\includegraphics{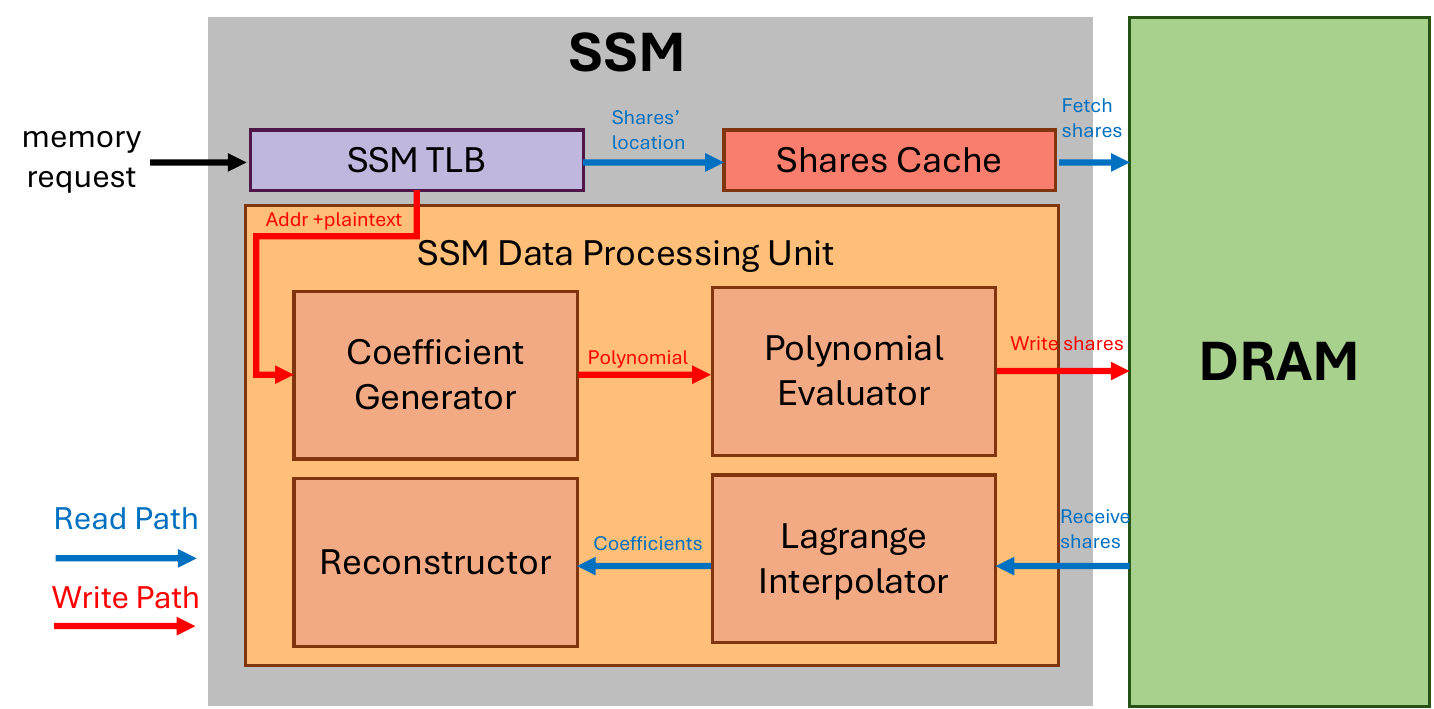}} 
 \vspace*{-2mm}
  \caption{SSM hardware architecture and data flow.}
  \label{fig:SSM_hardware}
  \vspace*{-5mm}
\end{figure}

\HPCA{Fig.\ref{fig:SSM_hardware} shows the hardware components required for SSM implementation, including the SSM TLB for address translation, shares cache for performance optimization, and the data processing unit containing four core cryptographic modules. The data processing unit comprises the coefficient generator to create polynomial coefficients from input data and seeds, the polynomial evaluator to generate shares by evaluating polynomials at random points, the reconstructor to extract coordinate pairs from retrieved shares, and the Lagrange interpolator to perform polynomial interpolation to recover original data.}

\HPCA{We implemented the four data processing unit components with $N=9$ (default setting) and cache line size = 64 bytes, using RTL and synthesized them with a 28nm commercial PDK using Genus synthesis targeting 1GHz operation. The complete data processing unit requires 0.0096\,mm$^2$ area and consumes 77.88\,mW power, with a critical path length of 1.05\,ns, indicating that the design can achieve higher frequencies if needed. In our Gem5 simulations, we model both data segmentation and reconstruction with 40-cycle latency, matching the AES encryption latency used in competing designs~\cite{emcc,rmcc}. This modeling choice reflects that computation latency is not the critical performance bottleneck in secure memory systems, where memory access patterns and metadata overhead dominate performance. For memory components (SSM TLB and shares cache), we used SRAM macros from a commercial 28nm memory compiler under standard conditions (0.9V, 25$^{\circ}$C). The SSM TLB and 128KB shares cache consume 0.260\,mm$^2$ and 206.65\,mW combined. \textbf{Total SSM hardware components consume 0.270\,mm$^2$ and 284.53\,mW at 28nm nodes.}}


\subsection{System-Level Simulation Infrastructure}

\begin{table}[t]
\centering
\caption{Simulation settings}
\label{tab:Sim_para}
\begin{tabular}{l|l}
\hline
\multicolumn{2}{c}{\textbf{Processor Parameters}}                    \\ \hline
Core            & 4 Core, X86, OoO, 3GHz                \\ \hline
L1 Cache        & 2 cycles, 32KB, 2-Way                      \\ \hline
L2 Cache        & 20 Cycles, 1MB, 8-Way                      \\ \hline
LLC             & 32 Cycles, 8MB, 16-Way                    \\ \hline
 TLB             & 2 cycles, 256 entries            \\ \hline

Page Size & 4 KB \\\hline
Page Walk Latency      & 4-level, each level 30 cycles   \\ \hline
\multicolumn{2}{c}{\textbf{Memory Parameters}}          \\ \hline
Type            & DDR4\_2400\_16x4                          \\ \hline
Size            & 32 GB                                       \\ \hline

\multicolumn{2}{c}{\textbf{AES-based Secure Memory Parameters}}      \\ \hline
AES Latency & 128-bit, 40 Cycles                                   \\ \hline
Authentication latency &  40 Cycles                                   \\ \hline
Merkle Tree      & 8 B per nodes, 8-way   \\ \hline
VN cache             & 128 KB, LRU \\ \hline
\multicolumn{2}{c}{\textbf{SSM Parameters}}      \\ \hline

Data Segmentation latency       & 40 Cycles  \\ \hline

Data Reconstruction latency       & 40 Cycles  \\ \hline

Shares Cache Size       &  128 KB (default), LRU  \\ \hline

SSM TLB      &  512 entries  \\ \hline
Share Size              & 9 B  \\   \hline
Access Overhead       &  8 (k=56 default)\\ \hline

\end{tabular}
\vspace*{-3mm}
\end{table}

To evaluate SSM, we implemented both SSM and Intel SGX within the Gem5 cycle-level simulator \cite{gem5}, utilizing its System Emulation (SE) mode. This SE mode allows for detailed architectural simulation without the overhead of a full operating system. The simulation setup is detailed in Table \ref{tab:Sim_para}, assuming a 4-core X86 architecture with 32GB DDR4 memory. \HPCA{Unlike SGXv1 systems, which are limited to only 128MB of secure memory, our setup assumes a 32GB secure memory configuration to expose the performance bottlenecks of SGXv1-like schemes and highlight the scalability and efficiency benefits of SSM under large secure memory allocations.} SGXv1-like was modeled with an 8-way MT, 64-bit VNs and MACs per 8 data blocks. We also assume an AES encryption process with a latency of 40 cycles and authentication latency of 40 cycles, using the same setup as in previous work \cite{emcc,rmcc,mgx,softvn}.

For fair comparison, we evaluated SSM alongside AES-based secure memory designs including AES-XTS, AES-GCM, and SGXv1-like systems. For AES-XTS and AES-GCM, we focused on performance overheads: AES-XTS incurs only AES latency without additional memory traffic, whereas AES-GCM introduces memory traffic overhead for VN management. For SGXv1-like designs, we compared seven representative optimizations: MorphCTR~\cite{saileshwar2018morphable}, EMCC~\cite{emcc}, RMCC~\cite{rmcc}, MGX~\cite{mgx}, SoftVN~\cite{softvn}, CTR-L~\cite{counterlight}, and COSMOS~\cite{cosmos} (see Sec.~\ref{sec:Experimental Results}).

%% file: 07_Experimental_Results.tex
\section{Experimental Results}
\label{sec:Experimental Results} 

This section verifies SSM's data segmentation and reconstruction via RTL simulation, analyzes performance overhead by examining key parameters' effects on execution time and cache hit rate, and compares SSM against other secure memory designs. For all execution time comparisons, we use normalized clock cycles, with the non-protected system as the baseline (normalized to 1).

\subsection{Data Reconstruction Accuracy in Hardware}

To validate the correctness of SSM’s datapath, we implemented the coefficient generator, polynomial evaluator, share generator, and Lagrange reconstruction modules in SystemVerilog and verified them using Cadence Xcelium RTL simulation. All coefficients are fixed to 8 bytes, and polynomial evaluation and interpolation are performed in $\mathrm{GF}(2^{64})$. The test uses a 64-byte cache line (eight 8-byte words), consistent with the SSM data path. As outlined in Sec.~\ref{sec:SSM_example}, a degree-9 polynomial is formed from eight data words and one coefficient seed. For degrees below 9, a single cache line is split into multiple polynomials; for degrees above 9, multiple cache lines are combined to form one polynomial. For each polynomial degree, the hardware generates shares for 100 random cache lines and reconstructs them via finite-field Lagrange interpolation, verifying correct polynomial generation and recovery.

Across all tested polynomial degrees from $N{=}2$ to $N{=}32$, the RTL simulation reports \emph{zero reconstruction errors}: every reconstructed cache line matches the original input exactly. This confirms that the SSM datapath preserves finite-field algebraic correctness without numerical or hardware-induced inaccuracies. Although the hardware can therefore support large degrees, doing so increases system-level overhead—higher degrees require more shares per data block, raising storage cost and memory traffic during reconstruction. At the same time, larger degrees strengthen security by requiring an adversary to obtain more genuine shares to recover the secret. 


\subsection{Polynomial Degree Choice}


\begin{figure} [t]
  \centering
 \resizebox{.95\columnwidth}{!}{\includegraphics{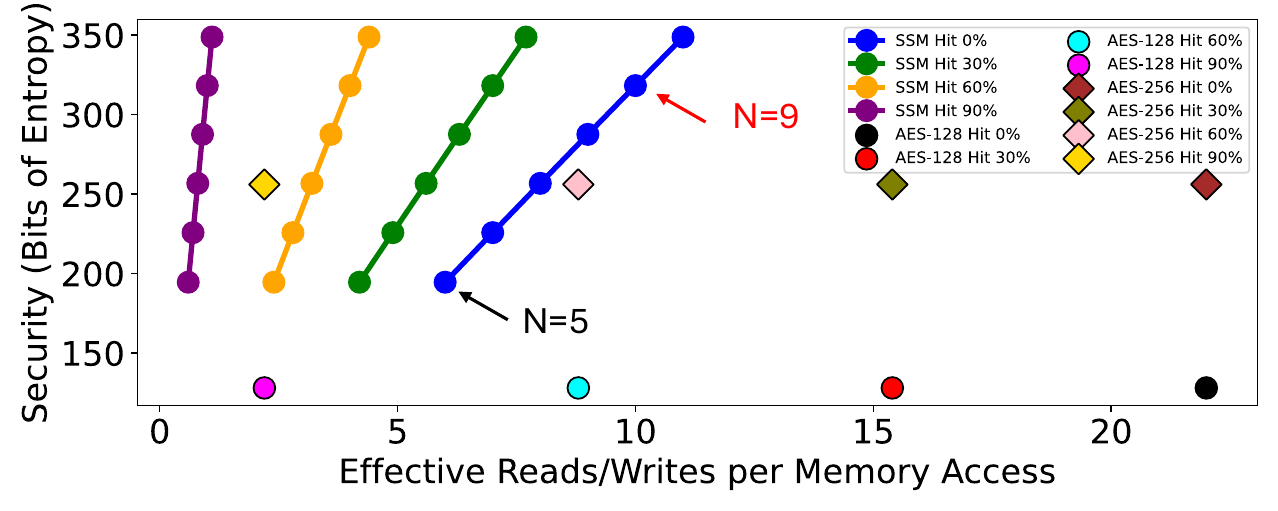}} 
 
  \caption{Security-performance trade-off between SSM and AES-based memory encryption across different cache hit rates, showing SSM's tunable security levels versus AES's fixed 128-bit security.}
  \label{fig:Mem_vs_security}
  \vspace*{-5mm}
\end{figure}

\HPCA{To determine the optimal polynomial degree for SSM, we analyze the security-performance trade-off for degrees ranging from 2 to 32. Higher polynomial degrees enhance security by requiring attackers to obtain more shares for data reconstruction, but this comes at the cost of increased memory accesses and performance overhead. Conversely, lower degrees improve performance through fewer memory accesses but reduce the security level. We quantify security using combinatorial entropy, calculated as $\log_2\binom{Total}{t}$, where $Total$ represents the total number of possible memory locations (for example when N=9, there are 28 billion shares for a 32GB memory) and $t$ is the number of shares an attacker must compromise. This metric represents the computational complexity of a brute-force attack attempting to locate all the required shares.}

\HPCA{Figure~\ref{fig:Mem_vs_security} illustrates this trade-off by comparing SSM's tunable security levels with AES-128's fixed 128-bit security and  AES-256's fixed 256-bit across different cache hit rates. The effective memory accesses per request are calculated as $(1 - \text{hit rate}) \times \text{number of shares}$ for SSM and $(1 - \text{hit rate}) \times \text{number of secure metadata access}$ for AES. The results show that with a 90\% cache hit rate, SSM achieves similar security to AES-256 at degree 9 while requiring only 0.9 effective memory accesses compared to AES's 2.2 accesses. For typical cache hit rates (60-90\%), polynomial degrees between 5-10 provide an optimal balance, delivering 100+ bits of security with significantly lower memory overhead than AES-based alternatives.}

\textbf{We select polynomial degree $N{=}9$ as the default SSM configuration}, as it balances security and efficiency. A degree-9 polynomial fits a 64-byte cache line naturally—eight 8-byte words plus one seed—simplifying the datapath while providing 256-bit equivalent security (AES-256). Our RTL results confirm full hardware feasibility with perfect reconstruction. For applications prioritizing performance or lower overhead, a smaller degree such as $N{=}5$ is also practical: a cache line can be split into two degree-5 polynomials (four words plus one seed), yielding 128-bit equivalent security (AES-128) with reduced overhead.

\begin{figure}
  \centering
\vspace*{-1mm}
 \resizebox{.98\columnwidth}{!}{\includegraphics{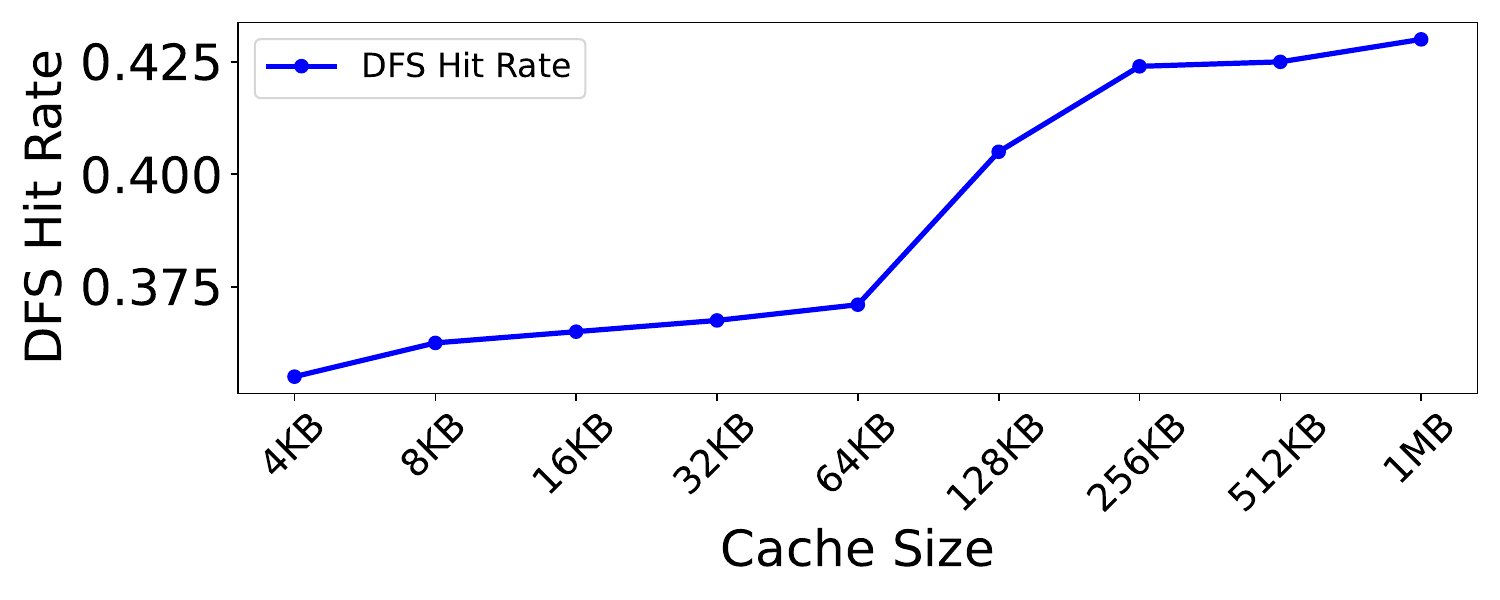}} 
 \vspace*{-3mm}
  \caption{SSM shares cache hit rate vs. size  for DFS.}
  \label{fig:shares_cache_size}
  \vspace*{-5mm}
\end{figure}

\subsection{Shares Cache Size Choice}
\HPCA{We mainly focus on evaluating the shares cache performance for irregular memory access patterns, as regular patterns already achieve high shares cache hit rates ($>$90\% at 64KB cache size). Specifically, we evaluate the system with fixed $N=9$ using DFS, an application exhibiting irregular memory access patterns. Fig.~\ref{fig:shares_cache_size} illustrates the cache hit rate as the shares cache size increases from 4KB to 1MB. For DFS, the hit rate improves from 36\% at 4KB to 43\% at 128KB, after which it plateaus with minimal further gains. Therefore, we select 128KB as the default shares cache size for synthesis and subsequent analysis, since it offers an optimal trade-off between hit rate improvement and hardware overhead—larger caches yield diminishing returns but significantly increase chip area and power consumption.}

\subsection{Application Level Evaluation}

\HPCA{We evaluate SSM's performance against three categories of AES-based secure memory systems: AES-XTS, AES-GCM, and SGXv1-like. For SSM, we use the optimized parameters determined from our analysis: polynomial degree N=9 and a 128KB shares cache. To ensure fair comparison, we equip the SGXv1-like system with a 128KB VN cache, matching SSM's cache capacity. We implement MorphCTR \cite{saileshwar2018morphable} as our SGXv1-like baseline, as it represents the most advanced algorithmic optimization in this category and serves as the foundation for subsequent improvements like EMCC\cite{emcc}, RMCC \cite{rmcc}, and CTR-L \cite{counterlight} Our evaluation employs two distinct workload categories: ML benchmarks that exhibit regular memory access patterns and graph algorithms that demonstrate irregular memory access patterns, using the same experimental setup described in Section~\ref{sec:Motivation}.}

\begin{figure} [t]
\centering
 \resizebox{0.98\columnwidth}{!}{\includegraphics{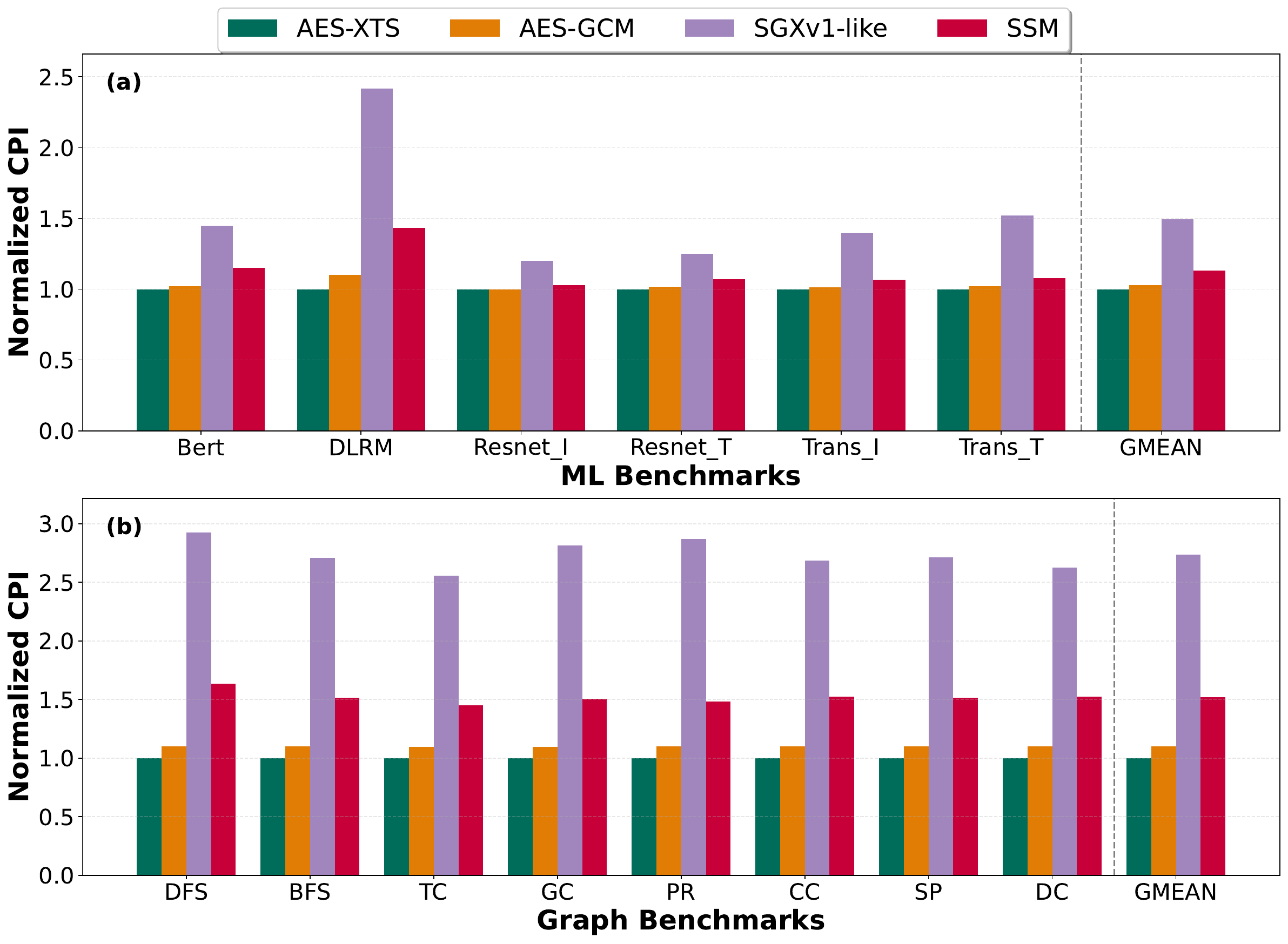}}
 \vspace*{-3mm}
\caption{Normalized CPI comparison of encryption schemes. (a) ML Benchmarks and (b) Graph Benchmarks. All values are normalized to AES-XTS, with lower values indicating better performance.}
\label{fig:CPI_comparison}
\vspace*{-5mm}
\end{figure}

\begin{figure} [t]
\centering
  \vspace*{-3mm}
 \resizebox{0.98\columnwidth}{!}{\includegraphics{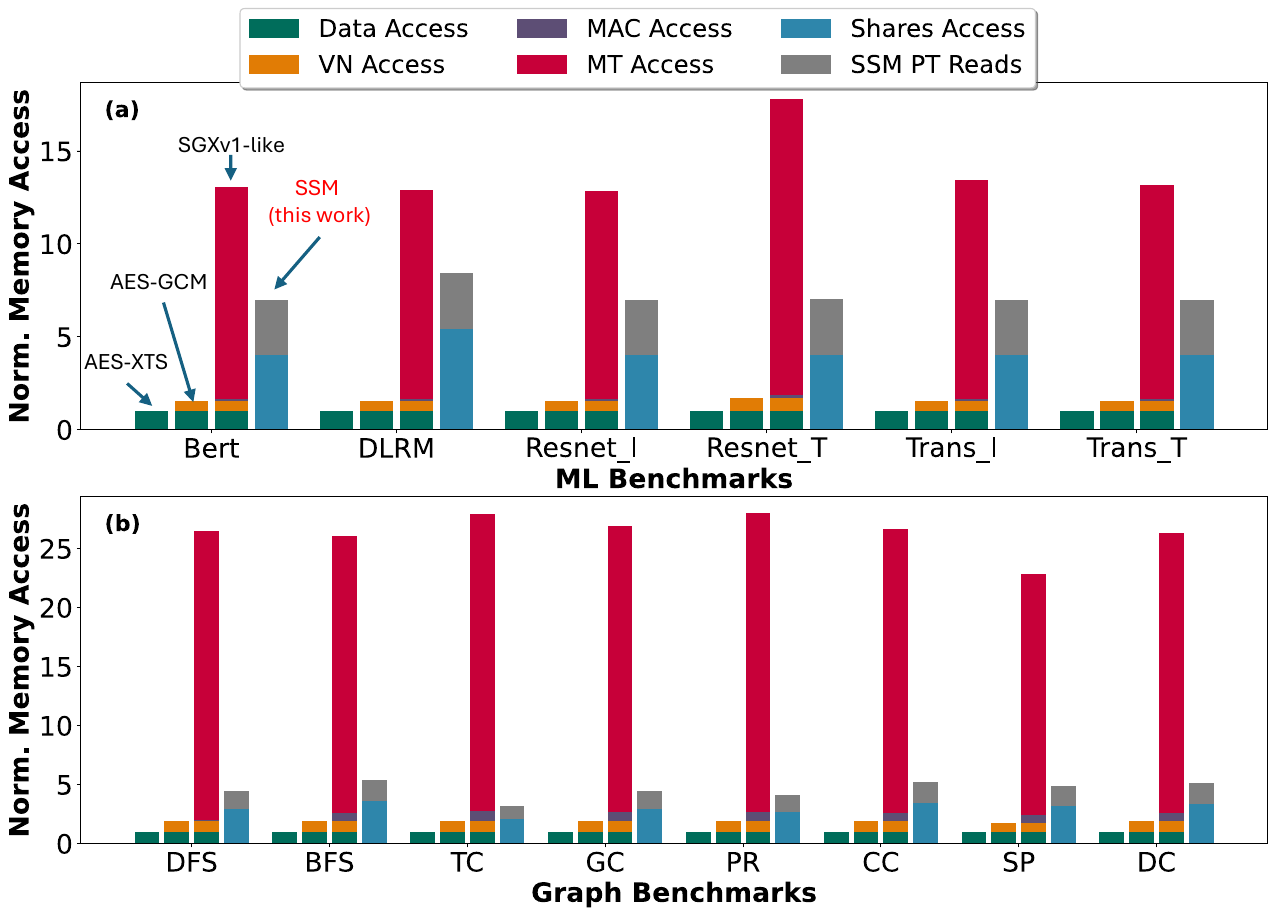}}
 \vspace*{-3mm}
\caption{Normalized memory access overhead of encryption schemes. (a) ML Benchmarks and (b) Graph Benchmarks. All values are normalized to data access.}
\label{fig:mmemory_traffic_SSM}
\vspace*{-2mm}
\end{figure}

\begin{figure} [t]
\centering
  \vspace*{-1mm}
 \resizebox{0.98\columnwidth}{!}{\includegraphics{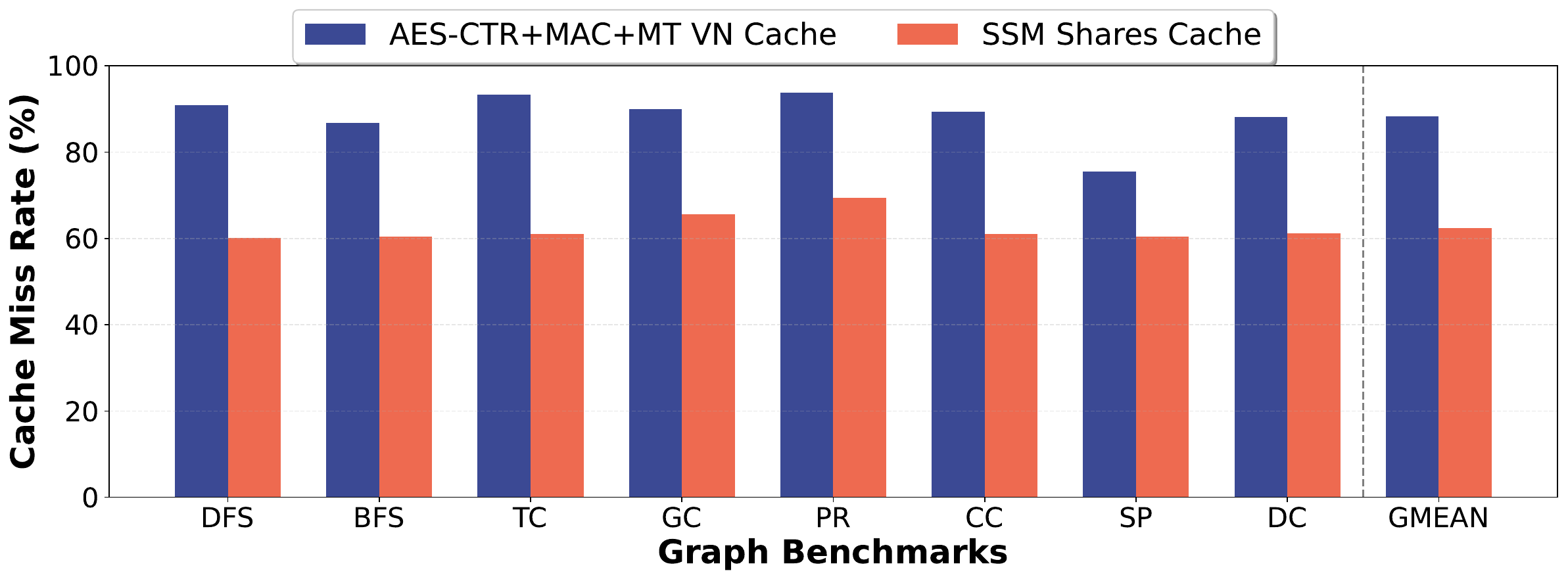}}
 \vspace*{-3mm}
\caption{Cache miss rate comparison between SGXv1-like VN Cache and SSM Shares Cache across graph workloads.}
\label{fig:cache_performance}
\vspace*{-7mm}
\end{figure}

\HPCA{Fig.~\ref{fig:CPI_comparison} shows the normalized CPI in four encryption schemes. Although SSM does not outperform AES-XTS or AES-GCM in terms of raw CPI—due to its stronger cryptographic protection—it provides the same level of security as SGXv1-like and achieves significantly better performance.  SSM is on average 12.5\% faster than SGXv1-like and in ML workloads and 43\% faster in graph workloads. The root cause of this performance gap is illustrated in Fig.~\ref{fig:mmemory_traffic_SSM}, which presents normalized memory accesses. Despite using MorphCTR, SGXv1-like still incurs high overhead from repeated MAC and MT node accesses—reaching over 25$\times$ normalized MT access in graph benchmarks. In contrast, SSM eliminates such deep tree traversals. Its primary sources of memory traffic stem from share accesses (including both reads and writes) and page table reads required for address translation. Together, these typically remain under 5.5$\times$ normalized access, resulting in substantially lower and more stable memory overhead across both ML and graph workloads.}

\HPCA{The performance benefits of SSM’s data-to-share mapping strategy are illustrated in Fig.~\ref{fig:cache_performance}, which compares the cache miss rates of SSM’s shares cache with the SGXv1-like VN cache. The figure highlights that SSM delivers significant improvements primarily in graph benchmarks, where its mapping strategy—consecutive addresses mapped to the same physical location with different share indices—enhances spatial locality. This is particularly effective for graph workloads, which often exhibit irregular and sparse memory access patterns. As a result, SSM reduces cache miss rates by an average of 30\% compared to SGXv1-like, with benchmarks like DFS and TC improving from over 90\% to around 60\%. In contrast, for ML benchmarks, both VN and shares caches already exhibit low miss rates (around 5\%), leaving little room for further optimization. This contrast highlights that SSM’s mapping strategy is especially beneficial in irregular workloads such as graph processing, where traditional caching mechanisms are less effective.}

\subsection{Comparison with Prior Works}
To provide a comprehensive evaluation, we compare SSM against recent state-of-the-art secure memory systems, including EMCC \cite{emcc}, RMCC \cite{rmcc}, COSMOS \cite{cosmos}, CTR-L \cite{counterlight}, MGX~\cite{mgx}, and SoftVN~\cite{softvn}. We implement an EMCC-like design in Gem5 by placing the counter cache at the L2 level, allowing counter accesses to proceed in parallel with L2, LLC, and DRAM data accesses. Since RMCC reports performance similar to that of EMCC over MorphCTR, we use our EMCC results as a proxy for RMCC. For COSMOS, we use the reported numbers from its original publication. The performance of CTR-L is derived from Figure 16 of its original publication \footnote{We use the performance of CTR-L with AES-128, while our SSM with $N=9$ offers AES-256 equivalent security. Thus, we compare SSM’s highest-security mode against CTR-L’s faster, lower-security setting.}. For MGX and SoftVN, we use both performance and design complexity data as reported in their respective publications. As EMCC, RMCC, and CTR-L primarily target graph workloads, we compare SSM with them using our graph benchmark suite. In contrast, MGX and SoftVN focus on ML workloads, so we compare their reported results with the performance of SSM on the ML benchmarks.

\begin{figure}
\centering
 \resizebox{0.98\columnwidth}{!}{\includegraphics{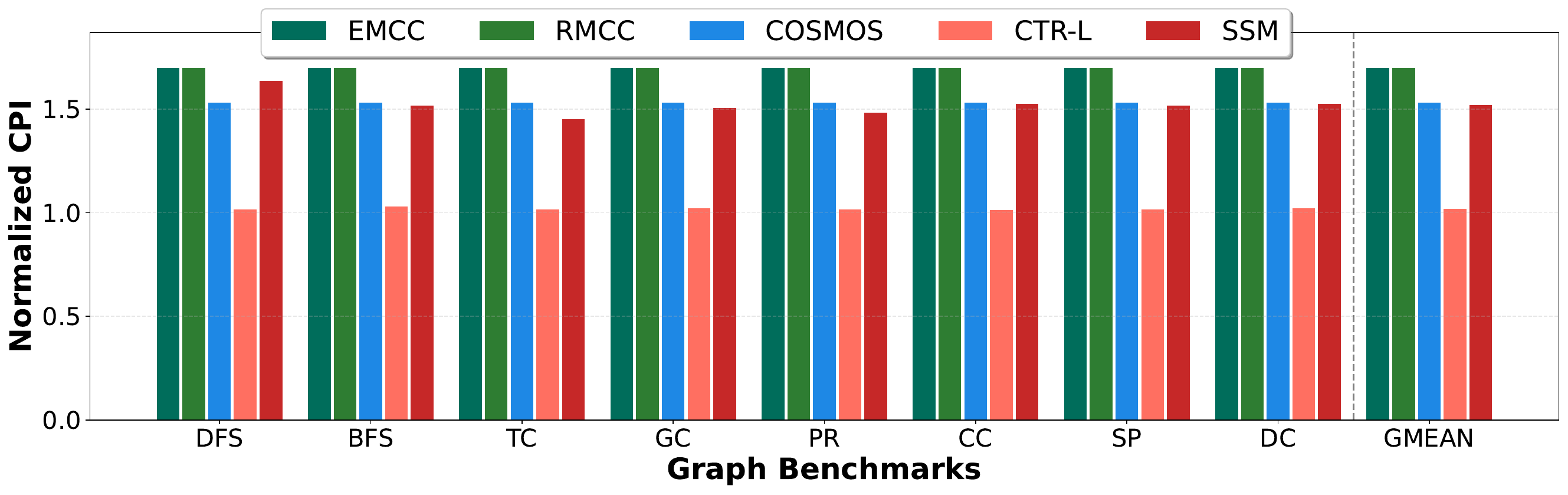}}
 \vspace*{-3mm}
\caption{Normalized CPI comparison of state-of-the-art secure memory designs on graph benchmarks. All values are normalized to non-protected memory.}
\label{fig:graph_compare_ssm}
 \vspace*{-5mm}
\end{figure}


Fig.~\ref{fig:graph_compare_ssm} shows the normalized CPI of secure memory designs on graph benchmarks, normalized to the non-protected baseline. SSM achieves about 12\% better performance than EMCC/RMCC and 3\% better than COSMOS. While EMCC, RMCC, and COSMOS rely on caching or prediction-based mechanisms that still incur counter metadata traffic, SSM eliminates counter accesses entirely through share regeneration, resulting in lower overhead and higher efficiency.

CTR-L, which combines AES-XTS and SGXv1-like encryption, achieves near-baseline performance with only 2\% overhead. Although SSM incurs about 15\% higher overhead, it delivers uniform protection across the entire memory space, preventing replay attacks through dynamic secret-share relocation. In contrast, CTR-L’s use of AES-XTS for hot data leaves those regions vulnerable to replay attacks~\cite{counterlight}. Thus, despite slightly higher cost, SSM offers stronger and more comprehensive security, making it well suited for high-assurance environments.


\begin{figure}
\centering
 \resizebox{0.98\columnwidth}{!}{\includegraphics{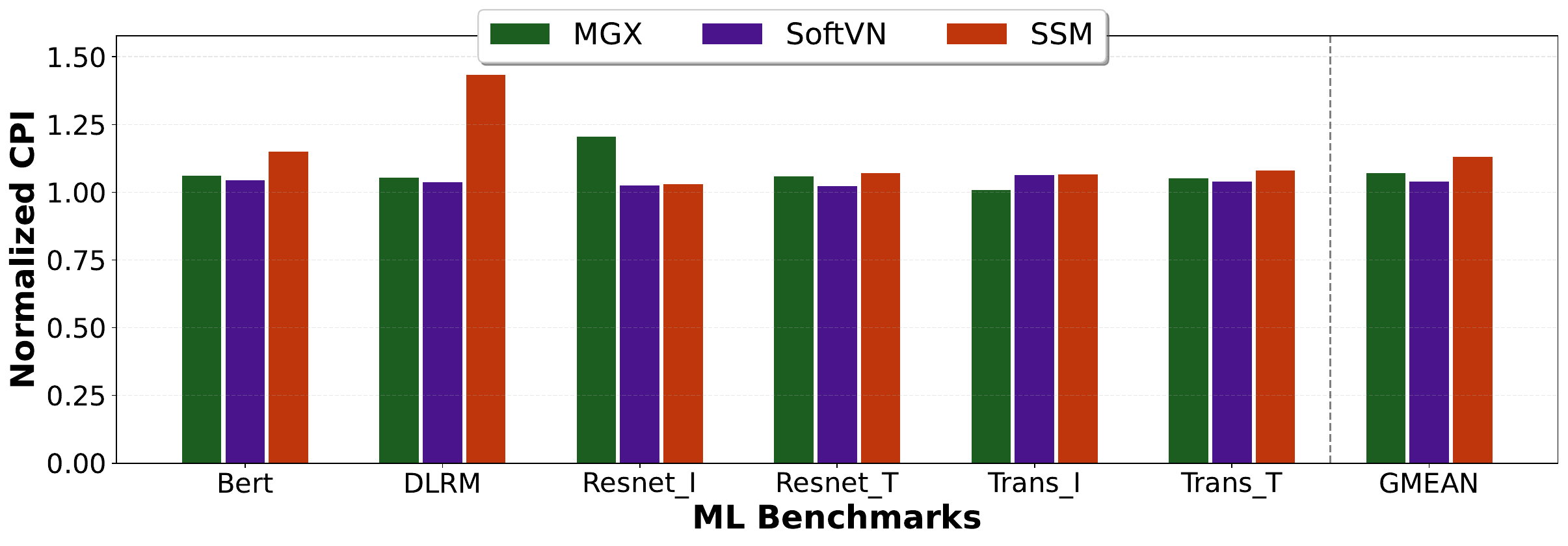}}
 \vspace*{-3mm}
\caption{Normalized CPI comparison of secure memory designs on ML benchmarks. All values are normalized to non-protected memory.}
\label{fig:ml_compare_ssm}
\vspace*{-5mm}
\end{figure}

Fig.~\ref{fig:ml_compare_ssm} shows normalized CPI on ML benchmarks. 
SSM incurs about 3\% higher overhead than MGX and SoftVN, which avoid off-chip VN storage. However, SSM offers key advantages: \circled{1} SSM functions as a general-purpose secure memory solution that works across diverse applications without requiring specialized hardware or software modifications. In contrast, MGX depends on dedicated hardware accelerators, while SoftVN requires specific software-hardware configurations \cite{softvn}. \circled{2} SSM provides stronger security guarantees. MGX's reliance on specialized accelerators creates potential vulnerabilities to side-channel attacks if an adversary compromises the accelerator. Similarly, SoftVN's software-generated VNs can be manipulated by attackers with system access. SSM's share-based approach operates independently of specific hardware or software components, delivering robust security without these architectural dependencies. Thus, SSM represents a more versatile and secure design that balances performance with comprehensive protection.

%% file: 08_Related_Work.tex
\section{Related Work} 

Extensive research on secure memory has led to counter-mode encryption schemes~\cite{SGX,suh2003efficient} with optimizations such as counter-based integrity trees~\cite{elbaz2007tec,hall2006parallelizable}, metadata caching~\cite{lee2016reducing}, and speculative VN usage~\cite{lehman2016poisonivy,shi2005high}. To reduce off-chip VN storage, MGX~\cite{mgx} and SoftVN~\cite{softvn} employ application-specific customization and software-generated VNs. More recent performance-oriented designs include EMCC~\cite{emcc}, which caches VNs in L2 for parallel access; RMCC~\cite{rmcc}, which memoizes cryptographic operations; COSMOS~\cite{cosmos}, which applies reinforcement learning to predict and prefetch counters; and Counter-light Encryption~\cite{counterlight}, which selectively uses counterless encryption. However, EMCC and RMCC retain substantial storage overhead, COSMOS still incurs metadata traffic from counter management, and Counter-light’s partial encryption approach remains vulnerable to replay attacks.

%% file: 09_Conclusion.tex
\section{Conclusion and Future Work}
We present Secure Scattered Memory (SSM), an efficient approach to memory security without using AES-CTR.  SSM efficiently protects data content and facilitates integrity checks by dispersing data into secret shares. Experimental results show SSM incurs only 10\% and 8\% overhead versus AES-XTS and AES-GCM respectively, while achieving 40\% better performance than Morphable Counter and 12\% better than EMCC/RMCC and 3\% better than COSMOS—all with minimal hardware cost (0.270 mm², 284.53 mW at 28nm).

\textbf{Future work:} We plan to develop formal mathematical proofs of SSM’s  with  and further optimize the design toward zero-overhead secure memory with comprehensive protection.